# Titanium Silicide Islands on Atomically Clean Si(100): Identifying Single Electron Tunneling Effects

J.L. Tedesco, J.E. Rowe
*Department of Physics, North Carolina State University, Raleigh, North Carolina 27695-8202, USA*

R.J. Nemanich
*Department of Physics, Arizona State University, Tempe, Arizona 85287-1504, USA*

## Abstract

Titanium silicide islands have been formed by the ultrahigh vacuum deposition of thin films of titanium (< 2 nm) on atomically clean Si(100) substrates followed by annealing to ~800ºC. Scanning tunneling microscopy (STM) and scanning tunneling spectroscopy have been performed on these islands to record current-voltage (I-V) curves. Because each island forms a double barrier tunnel junction (DBTJ) structure with the STM tip and the substrate, they would be expected to exhibit single electron tunneling (SET) according to the orthodox model of SET. Some of the islands formed are small enough (diameter < 10 nm) to exhibit SET at room temperature and evidence of SET has been identified in some of the I-V curves recorded from these small islands. Those curves are analyzed within the framework of the orthodox model and are found to be consistent with that model, except for slight discrepancies of the shape of the I-V curves at current steps. However, most islands that were expected to exhibit SET did not do so, and the reasons for the absence of observable SET are evaluated. The most likely reasons for the absence of SET are determined to be a wide depletion region in the substrate and Schottky barrier lowering due to Fermi level pinning by surface states of the clean silicon near the islands. The results establish that although the Schottky barrier can act as an effective tunnel junction in a DBTJ structure, the islands may be unreliable in future nanoelectronic devices. Therefore, methods are discussed to improve the reliability of future devices.

**I. Introduction**

Single electron tunneling (SET) has been proposed to form the basis of the next generation of electronic devices [1-2], and is expected to occur in double barrier tunnel junction (DBTJ) structures, where a conductor (such as a metal island) is placed between two tunneling barriers [3]. For SET to be observed in the metal island, the resistances of both tunnel junctions must be greater than the quantum of resistance, $R_T > h/e^2 \sim 25.8$ k$\Omega$, where $h$ is Planck's constant and $e$ is the electron charge. If either barrier does not satisfy this requirement, the tunneling time, $t_\tau$, will be too short for the electrons to be localized on the island. The charging energy, $E_c$, of the island must satisfy the relation $E_C > kT$, where k is Boltzmann's constant and T is the temperature. The charging energy is characterized as $e^2/2C$, where C is the capacitance of the island, and electron transport through the island will be suppressed if this energy is greater than kT. For the second criteria to be satisfied at room temperature, the diameter of the metallic island must be ~10 nm or less [4].

In DBTJ structures, each tunnel junction is described as a parallel combination of a resistor and a capacitor [3], as is shown in Fig. 1(a). When the two tunnel junctions are symmetric ($R_1C_1 = R_2C_2$), the Coulomb blockade is the only discernible single electron effect [4]. The Coulomb blockade occurs because charge quantization results in a gap of width $2e/C$ to form in the states available for tunneling [5]. Outside of this gap, the current-voltage (I-V) curve is ohmic because tunneling events occur simultaneously. However, when the tunnel barriers are asymmetric ($R_1C_1 > R_2C_2$ or vice versa), the current increases in a stepwise fashion at regular intervals as the bias voltage is increased [3], and this is known as the Coulomb staircase.

It is important to establish if the Schottky barrier may be used as one of the tunnel barriers for SET in metal island-semiconductor substrate systems. Metal islands are known to self-assemble on semiconductor substrates [6-9] and the ability to use the Schottky barrier as an effective tunnel junction could minimize the number of processing steps necessary to form nanodevices. Furthermore, nanodevices composed of metallic islands could be made much smaller than devices composed of semiconducting islands because the electron density in metals is much higher (n ~ $10^{15}$ cm$^{-3}$ to $10^{19}$ cm$^{-3}$ for semiconductors while n ~ $10^{22}$ cm$^{-3}$ for metals). The higher electron density allows small islands to have large charging energies and operate at temperatures above 4.2 K [10].

While SET has been proposed for several metal nanocluster systems, such as Au nanoclusters on organic self-assembled monolayers [11-12], metal particles encased in organic molecules [13-14], and Ag and Au islands on semiconducting surfaces [15-22], each of these materials systems would have difficulty withstanding the temperatures and processing necessary for device fabrication. Given the considerations of both functionality and stability, TiSi$_2$ on silicon may prove to be an ideal foundation for future nanoelectronic devices. TiSi$_2$ is already used in current integrated circuit technology [23], and the TiSi$_2$/Si system is capable of withstanding the high temperatures used in device fabrication. Furthermore, measurements from our laboratory and elsewhere have indicated SET characteristics at room temperature in self-assembled TiSi$_2$ islands on Si(111) [24-25]. Observations of SET in these islands are reasonable because small TiSi$_2$ islands on a silicon substrate would be expected to exhibit SET because the STM tip-TiSi$_2$ island-silicon substrate system is a DBTJ structure [4,24-25]. The vacuum gap at the tip-island interface is obviously a tunnel barrier and the Schottky barrier acts as a barrier to tunneling in these experiments because the electrons must overcome it in order to be transported across the interface [26]. However, an additional study in our laboratory has shown that the Schottky barrier can be influenced by surface states of the clean silicon surface [27], which could potentially affect the ability of the TiSi$_2$ island system to exhibit SET.

Previously, SET was most prominently observed after the $TiSi_2$ islands were formed on an epitaxial layer of intrinsic silicon [24]. In the current study, $TiSi_2$ islands were formed on atomically clean Si(100) without an intrinsic layer and current-voltage (I-V) and differential conductance ((dI/dV)-V) curves of the islands were recorded in ultrahigh vacuum (UHV) conditions. In the absence of the layer of intrinsic silicon, the Schottky barrier at the island-substrate interface and island edges play a more prominent role in the transport properties. A simplified drawing of the tip-island-substrate structure is shown in Fig. 1(b). The tip-island interface corresponds to the leftmost RC circuit ($R_1$, $C_1$) shown in Fig. 1(a), while the island-substrate interface corresponds to the rightmost RC circuit ($R_2$, $C_2$). Figure 1(c) shows the band structure of the tip-island-substrate DBTJ from Fig. 1(b). The band structure shows the zero bias case for a tungsten STM tip [26,28], a $TiSi_2$ island with a Schottky barrier height of 0.6 eV [28], and the band bending, Fermi level, and depletion width consistent with an n-type silicon substrate doped at $1 \times 10^{17}$ $cm^{-3}$ [28-29]. While a previous study reported that the barrier heights of $TiSi_2$ islands vary for nanoscale islands [30], that same study reported no correlation between island size and barrier height. Therefore, because 0.6 eV is the accepted bulk value for $TiSi_2$ [28], it is used as a convenient example of the barrier height for the representative band structure shown in Fig. 1(c). The I-V and (dI/dV)-V curves that appear to exhibit SET signatures are compared with theoretical predictions in order to establish that the SET is genuine. Evidence of genuine SET would suggest that it is possible to use a Schottky barrier alone as one of the tunnel barriers in SET-based nanoelectronic devices.

## II. Experimental Procedure

*II.A. Formation of TiSi$_2$ islands and the measurement of SET*

The experiments were performed with a commercially-available surface science system (Omicron Nanotechnology Multiprobe P) consisting of a preparation chamber and an analysis chamber. The preparation chamber is equipped with a triple-cell electron beam evaporator (Focus EFM 3T) for *in situ* deposition. The analysis chamber is equipped with a variable temperature UHV STM (Omicron Nanotechnology VT AFM), and has low energy electron diffraction (LEED) and Auger electron spectroscopy (AES) capabilities. The base pressure in the preparation chamber was $\sim 7.0 \times 10^{-11}$ torr, and the base pressure in the analysis chamber was $\sim 1.5 \times 10^{-11}$ torr.

Substrates were cut from 25.4 mm diameter, n-type (phosphorus-doped) silicon wafers with thicknesses of 0.250 mm ± 0.025 mm and doping concentrations of $\sim 8.0 \times 10^{16}$ to $\sim 2.0 \times 10^{17}$ cm$^{-3}$ (as determined from the resistivities, $\rho = 0.05$ to $0.10$ Ω-cm). A diamond-tipped scribe was used to section the wafers into strips $\sim 2 \times 10$ mm$^2$. These wafer sections were mounted on to sample cartridges and loaded into UHV without an *ex situ* chemical clean. To avoid significant outgassing during sample heating, the sample cartridge was degassed for several hours in UHV at ~500°C. After this initial heat treatment, direct current heating was used to hold the sample at ~650°C for ~12 hours (typically overnight). A clean surface was then obtained by flashing the sample in 5 second increments at steadily increasing temperatures, culminating with 2 to 4 flashes at ~1150°C, each lasting 30 seconds. The temperature was measured using an Ultimax optical pyrometer with the emissivity, ε, set at 0.65 [31]. During flashing, the pressure remained below $1.5 \times 10^{-9}$ torr. Following flashing, LEED was used to confirm the Si(100):2×1 reconstruction. After observing the reconstructed LEED pattern, STM and scanning tunneling spectroscopy (STS) were used to characterize the state of the surface. The surface was considered suitable for island formation if it was clean, flat, nominally free of defects, and showed atomic terraces with widths > 10 nm. The STM and STS measurements were performed using electrochemically etched tungsten tips.

Once the surfaces were confirmed to be clean and flat, the samples were transferred to the preparation chamber for deposition. The deposition source was a 2 mm diameter titanium rod of 99.99% purity (Goodfellow). Titanium layers, 0.1 nm to 0.2 nm thick, were deposited with the sample at room temperature. The deposition rate was calibrated by depositing a thick layer of titanium while maintaining a constant flux and then measuring the resulting layer thickness using an ambient atomic force microscope (AFM: Park Scientific Instruments Autoprobe M5). During deposition, the pressure in the preparation chamber did not rise above $1.5 \times 10^{-9}$ torr and was generally lower. After deposition, AES was used to confirm the presence of titanium and then samples were annealed at ~800°C for 60 seconds. Following annealing, LEED measurements were performed. If a diffraction pattern was detected, it was presumed that the titanium had formed into $TiSi_2$ islands, exposing the underlying silicon substrate. Samples were then transferred into the STM for scanning and STS measurements. The STM scans were recorded with the tip biased between +1.0 V and +2.5 V and a tunneling setpoint of between 0.75 nA and 1.0 nA. The I-V curves were recorded from -2.5 V to +2.5 V and were numerically differentiated to obtain the (dI/dV)-V curves.

*II.B. Analytical model*

According to the orthodox model of SET [32-35], the width of the steps in the Coulomb staircase should correspond to $\Delta V = e/C_\Sigma$, where $C_\Sigma = C_I + C_T$, where $C_I$ is the island-substrate capacitance and $C_T$ is the tip-island capacitance [36-38]. The island-substrate capacitance, $C_I$, can be approximated by considering the nanostructure to be a metal sphere of diameter d surrounded by a material with dielectric constant κ, $C_I = 2\pi\kappa\varepsilon_0 d$, where $\varepsilon_0$ is the permittivity of free space. A common approximation is to locate the sphere in a vacuum (κ = 1) [4,24,39-40], which allows for evaluation of the self-capacitance of the island using a simple model. The approximation for $C_T$ assumes that the radius of the STM tip is much less than the radius of the island; therefore the island is a semi-infinite plane compared to the tip atom (in this study, the smallest islands are over 20 times larger than the atoms of the STM tip). Thus, the tip-island system can be modeled as a sphere of radius r separated by a distance *s* in a material of dielectric constant κ. Assuming higher order terms are negligible, $C_T \approx 2\pi\kappa\varepsilon_0 r[\ln(s/r)+\ln(2)+(23/30)]$ [41]. Utilizing ΔV and the tunneling spectra from the islands, the $C_\Sigma$ values for the islands can be determined and compared to the theoretical values predicted using the equations for $C_T$ and $C_I$.

## III. Results

Figure 2(a) shows TiSi$_2$ islands grown on an n-type Si(100) surface and imaged at room temperature. Current-voltage spectra were recorded from several islands in the scan area. The islands in Fig. 2(a) labeled "2" and "5" have diameters of 7.0 ± 0.3 nm and 3.0 ± 0.3 nm, respectively, as shown in the line scans shown in Figs. 2(b) and 2(c). Therefore, both islands are small enough to expect to observe SET at room temperature. Figure 2(d) shows a linescan of the island labeled "8," which has a diameter of 20.0 ± 0.5 nm. Island "8" is not expected to exhibit SET at room temperature, and serves as a control to ensure that I-V curves are artifact-free. As shown in Fig. 3, the I-V curves recorded from islands "2" and "5" demonstrate a series of regular steps and the (dI/dV)-V curves exhibits regular peaks, while the I-V curve recorded from island "8" shows neither feature. Peaks in the (dI/dV)-V curves indicate incidents of increased conduction, which is a potential sign of SET.

Tunneling spectra were recorded from 261 round islands, such as those shown in Fig. 2(a). The size distribution of these islands is displayed in Fig. 4. As shown in Fig. 4, most of the islands are large and would not be expected to exhibit SET, however, ~49% of the islands had diameters less than 10 nm. Of these smaller islands, only the 2 islands identified in Fig. 2(a) exhibited step-like structures in their I-V curves reminiscent of SET. The fact that ~2% of the eligible islands exhibited evidence of SET is significant, and will be discussed in more detail below.

Current-voltage spectra were recorded from more than 261 islands; however, those islands that were not round in shape were excluded from the histogram in Fig. 4 due to the difficulty of determining reasonable approximations to use for calculating their self-capacitances. However, none of the other islands measured in this study exhibited SET.

**IV. Discussion**

*IV.A. Analysis and interpretation using the orthodox model*

Prior to concluding that the TiSi$_2$ islands exhibit SET, it is necessary to establish that the step-like I-V curves shown in Figs. 3(a) and 3(b) are due to SET and not artifacts. The steps in the I-V curves are equally spaced and the step spacings are the same for both positive and negative bias. Therefore, the step-like structures are not due to intra-band tunneling between level spacings because such tunneling would lead to step spacings that were different at positive and negative bias [14,42-43]. The step-like structures are also not due to the band gap and surface states of silicon because the position of the peaks in the differential conductance curves do not correspond to the positions of the surface states of silicon [44]. To eliminate tip contamination as the source of the step-like features, I-V curves were recorded on islands near those islands exhibiting step-like features. One such I-V curve is shown in Fig. 3(c) and does not exhibit regular step-like structures. Current oscillations, the oscillations caused by the partial reflection and interference of the electron wave as it tunnels [45], are not likely to be the source of the step-like structure in the I-V curves. Larger islands shown in Fig. 2(a) do not exhibit step-like structures in I-V curves recorded on them, as shown in Fig. 3(c), suggesting that current oscillations are not occurring. Furthermore, current oscillations are known to increase proportionally with increases in the current [22] and this was not observed in Figs. 3(a) and 3(b).

Therefore, the step-like I-V curves are attributed to SET and not artifacts, thus, the SET signatures can be identified and analyzed in detail.

The tip diameter $d_1$ and the tip-sample separation $s$ will be constant for the calculations of the $\Delta V$ values for both islands. Therefore, for these calculations, the tip diameter $d_1$ is 0.282 nm and the distance $s$ is 1 nm. While neither value was explicitly determined during the experiments, these approximate values are reasonable. When a tunneling tip is capable of atomic resolution, the tip can be approximated as spherical [46], the diameter of which would be $d_1$. Given that the radius of a tungsten atom is ~0.141 nm [47], $d_1$ is reasonable. Furthermore, the approximation of $s$ is reasonable given that the tunneling setpoint is 1 nA. From the tunneling spectra shown in Fig. 3(a), $\Delta V$ for the island labeled "2" in Fig. 2(a) is $0.33 \pm 0.02$ V. Therefore, the $C_\Sigma$ value for this island is expected to be $4.85 \times 10^{-19} \pm 0.29 \times 10^{-19}$ F. From the line scan shown in Fig. 2(b), the diameter of island "2" is $7.0 \pm 0.3$ nm. Using this value as $d_2$, $C_\Sigma \approx 4.14 \times 10^{-19} \pm 0.01 \times 10^{-19}$ F. Using this value of $C_\Sigma$, the predicted value of $\Delta V$ is $0.38 \pm 0.02$ eV. For the island labeled "5" in Fig. 2(a), $\Delta V$ is ~$0.63 \pm 0.02$ V from the tunneling spectra shown in Fig. 3(b). Therefore, the $C_\Sigma$ value for island "5" is expected to be ~$2.54 \times 10^{-19} \pm 0.09 \times 10^{-19}$ F. From the STM line scan shown in Fig. 2(c), the diameter of the indicated nanostructure is determined to be $3.0 \pm 0.3$ nm. Using the same process as above, $C_\Sigma \approx 1.96 \times 10^{-19} \pm 0.02 \times 10^{-19}$ F and the predicted value of $\Delta V$ is $0.82 \pm 0.07$ eV.

The approximations used to calculate the above values of $C_\Sigma$ and $\Delta V$ are the simplest. It is likely that more realistic models would yield different results. However, the models used in the above calculations provide lower bounds to the theoretical values. Given that the islands are known to not be spherical in general, were the islands modeled as disks instead of spheres [36,48], the calculated capacitances would be ~1.5 times smaller than the above values and the calculated voltage spacings would be ~1.5 times larger than the above values. However, the qualitative results of the comparisons are independent of the choice of model for the DBTJ system. The agreement between the theoretical and measured values of $C_\Sigma$ and $\Delta V$ is not perfect, but it is within a reasonable margin of error given the approximations that were made. Furthermore, the fact that the differences between the theoretical and measured values for island "5" are greater than for island "2" is also reasonable given that island "5" is the smaller island. Therefore, the agreement suggests that the Coulomb staircases shown in Figs. 3(a) and 3(b) are genuine. Furthermore, the conductance increases shown in Figs. 3(a) and 3(b) are more pronounced for positive voltages than for negative voltages, which is consistent with SET observed from islands on n-type substrates [20].

*IV.B. Discrepancies with the predictions of the orthodox model*

There are several features in the recorded I-V curves exhibiting SET that differ from the predictions of the orthodox model. These features include: uneven step heights, asymmetric current increases at positive and negative voltages, and rounding of the top edge of the steps. In addition to equidistant steps in the I-V curve ($\Delta V$), the orthodox model predicts step heights that are equivalent (except for the first step, which is half the height of subsequent steps [3]). Due to these equal step heights, the heights of successive peaks in the (dI/dV)-V curve are also predicted to be equal. However, in the recorded I-V curves, the peaks are not equal, and their intensity is more pronounced on one side of zero bias than on the other. Each of the inconsistent features described can be explained by the fact that the tunneling processes in this study are more complicated than those considered in the original orthodox model.

In the STM tip-metal island-semiconductor substrate DBTJ system studied here, current transport occurs via thermionic emission over the Schottky barrier at the $TiSi_2/Si$ interface [26]. This current transport is non-linear because the effective barrier height is voltage-dependent and is based on the width of the depletion region. The depletion width varies due to the amount of band bending as the interface is biased [49]. Furthermore, while the barrier to tunneling decreases with increasing voltage when the interface is forward-biased, the barrier height remains nearly constant when the interface is reverse-biased [24]. Both facts explain the uneven step heights and the asymmetry in differential conductance peak heights between positive and negative voltage. The step heights are uneven because non-linear changes in the effective barrier height cause non-linear increases in current between tunneling events. Furthermore, such non-linear increases in the effective barrier height lead to differential conductance peaks that are more pronounced when the interface is forward-biased. For n-type substrates, such as those used in this study, the interface is forward-biased when the bias applied to the tip is positive, which explains why the peaks in the (dI/dV)-V curves shown in Figs. 3(a) and 3(b) are more pronounced for positive tip voltage than for negative tip voltage. The voltage-based asymmetry in barrier height changes is also the source of the asymmetry in the positive and negative current increases. As shown in Figs. 3(a) and 3(b), the value of the current at positive bias is greater than the value of the current at negative bias. This asymmetry is due to the fact that as the barrier height decreases with increasing positive bias, the tunneling rate increases, and consequently the current increases. As stated, the barrier height does not change significantly with increasing negative bias, leading to a tunneling rate that does not change significantly, and a current that does not increase as quickly with increasing voltage.

    The only non-orthodox feature not explained by the voltage-dependent effective barrier height at the $TiSi_2$/Si interface is the rounding of the steps in the I-V curves. According to the orthodox model, the increases in current occur when the applied voltage supplies enough energy to exceed the charging energy of an island. The electron tunnels through the barrier onto the island in the characteristic time $t_\tau$, which is on the order of $10^{-15}$ s [50]. Such short time intervals lead to sharp step edges in the I-V curve. In this study, however, the edges of the steps appear

rounded, as shown in Figs. 3(a) and 3(b). This rounding is likely due to thermal activation of the tunneling processes [51]. As the temperature is increased, states above the Fermi energy are populated and states below the Fermi energy are emptied [3]. As more states appear below the Fermi level, there are more states available for electrons to tunnel into, leading to a gradual rounding of the top edge of the steps as the temperature increases [50]. Furthermore, the steps naturally become dull as the bias increases [22], independent of temperature.

*IV.C. Possible explanations for the absence of SET in many islands*

As stated previously, ~2% of the islands with diameters less than 10 nm exhibited SET. Previous studies [20,22] have implied that the observation of SET in $TiSi_2$ islands is straightforward; however, this is evidently not true. Therefore, the significant absence of SET must be explained.

Several explanations for the lack of SET have been given in the literature that can be shown to not apply to this study. The suggestion that the quantum electromagnetic fluctuations could be preventing the occurrence of SET effects [52] is invalid because such electromagnetic fluctuations would not wash out SET in a DBTJ system [53]. An alternate explanation to the lack of SET is suggested by other studies that claim that SET can only occur in islands on semiconductor surfaces if the islands are very close together [17,54]. Those studies suggested that lateral electric conduction through a common space charge region was the origin of SET. However, in those studies, the islands were within a few angstroms of each other, and that is not the case in the present study nor was it the case in a previous study [24]. Furthermore, some islands formed close together, as shown in Fig. 5(a), but did not exhibit SET, as shown in Figs. 5(b) to 5(d). Therefore, the lack of lateral conduction is not the explanation for the lack of SET in those islands from which it would be expected.

The resistances of both tunnel barriers must be examined to determine if they exceed 25.8 kΩ. The resistance of the vacuum gap is ~200 to 800 MΩ, as determined by applying Ohm's Law to the values of the current at I > +1.5 V in the recorded I-V curves. If the resistance of the vacuum gap changed significantly, that would cause a significant increase in the current at high voltage in the I-V curves, which is not observed. However, for substrates with doping concentrations of ~$10^{17}$ cm$^{-3}$, the junction resistance at the TiSi$_2$/Si interface is ~50 to 100 kΩ [55], which is similar in magnitude to the quantum of resistance. The Schottky barrier height of nanoscale islands has been shown to be lowered significantly [27,30] relative to the bulk barrier height. Additionally, Fermi level pinning by surface states of the non-passivated surface can act to lower the barrier height [27,56]. There are additional mechanisms that could cause barrier lowering. Image force lowering is one such mechanism [30]. In substrates where n ~ $10^{17}$ cm$^{-3}$, image force lowering of the barrier accounts for ΔΦ ~ 0.03 eV [57]. Therefore, image force lowering of the Schottky barrier would be a minimal, but non-negligible, effect in this study. Local field enhancement due to interfacial faceting of individual islands has also been identified as a barrier height lowering mechanism [30]. If local field enhancement were occurring, the increased field density would lead to thermionic field emission and Schottky barrier lowering [28]. However, a study of macroscopic TiSi$_2$/Si contacts suggested that the barrier lowering effect due to thermionic field emission decreases with TiSi$_2$ thickness [58]. As shown in Fig. 2, the thickness of these islands (even assuming significant penetration of the islands into the substrate during formation) is only a few nanometers. Therefore, while barrier lowering due to field enhancement cannot be conclusively ruled out, it is likely a negligible effect. Due to the numerous mechanisms by which the Schottky barrier height of the TiSi$_2$ islands is lowered relative to the bulk value, the junction resistance of the TiSi$_2$/Si interface is likely equivalent to or less than 25.8 kΩ. In that case, SET would not be observed. It should be noted that several previous studies reporting SET from metal islands were performed using passivated substrates [12,16-19,21-22], lending credence to the theory that Fermi level pinning by the surface states of clean surfaces inhibits SET. The fact that SET is observed in a few islands suggests that for those islands the Schottky barriers are not pinned and further investigation is necessary to determine the reason.

One aspect of this DBTJ system that cannot be discounted is the effect of the depletion width. While the depletion width is not typically discussed in the orthodox model [32-35], for a DBTJ system featuring a metal-semiconductor interface, understanding the depletion width is critical to understanding the performance of the system. For SET to be observable, the depletion width must not be too wide or the steps in the I-V spectra will not be observable [59]. The depletion width, W, can be calculated using $W = (\varepsilon\varepsilon_0 V_{bb}/eN_d)^{1/2}$, where $\varepsilon$ is the dielectric constant of the substrate, $\varepsilon_0$ is the permittivity of free space, $V_{bb}$ is the band bending at the interface, and $N_d$ is the doping concentration of the substrate [29]. For the substrates used in this study, the depletion width can be calculated to be ~50 ± 10 nm. Even accounting for reduced band bending due to barrier lowering caused by the surface states of the clean surface [27], the depletion width would still be ~31 ± 9 nm. In previous studies involving junctions between metals and clean semiconductor substrates, the second tunnel barrier was significantly narrower than even 22 nm [24,59-60], suggesting that the depletion width is too wide in this study.

With a depletion width that is too wide to typically allow for observable SET, it is evident that another effect is at work that would allow for the small islands shown in Fig. 2(a) to exhibit SET. Assuming that the resistance of the $TiSi_2$/Si(100) junction exceeds 25.8 kΩ, the SET shown in Figs. 3(a) and 3(b) suggests that the depletion width has been narrowed for those islands. One mechanism that could cause the depletion width to decrease is an accumulation of titanium atoms in the vicinity of the islands. In previous studies [27,61], there was evidence that the clean silicon surface was influenced by metal impurities from the electron beam deposition. Given that the deposition technique was the same here, it is reasonable that titanium atoms could have accumulated on the "clean" surfaces in this study, leading to changes in the pinning of the surface Fermi level [27] as well as the zero bias conductance [61]. Furthermore, it has been suggested that the transport for nanoscale Schottky contacts, such as these $TiSi_2$ islands, would be dominated by a surface recombination-generation minority current along an electrically active surface surrounding the contacts [61]. Therefore, the titanium atoms could make the "clean" silicon surface electrically active, leading to a surface current that dominated majority carrier transport across the Schottky barrier, preventing SET from occurring. There is evidence of dimer rows in Fig. 2(a), implying that those areas are clear of significant accumulations of titanium atoms. Therefore, the surface surrounding those islands is electrically inactive, allowing majority current transport to dominate and SET to occur.

The Schottky barrier height lowering due to Fermi level pinning can be utilized in schemes to improve the reliability of these $TiSi_2$ islands in future nanoelectronic devices. Using surface or defect engineering, schemes can be devised where the Fermi level pinning of the surface is used to tune the Schottky barrier of the islands. Such tuning could be used to activate and deactivate the single electron tunneling properties of the islands, allowing an island-based nanoelectronic device to be turned on and off. Furthermore, surface engineering would enable the Schottky barrier to be tuned without changing the composition of either the islands or the device.

**V. Conclusion**

$TiSi_2$ islands have been formed on atomically clean Si(100). Current-voltage curves recorded on these islands at room temperature have shown evidence for SET and the steps in the Coulomb staircases were analyzed and were found to agree with the orthodox model. There were discrepancies between the shape of the recorded I-V curves and those predicted by the model, however, these were attributed to non-linear current transport over the Schottky barrier at the TiSi$_2$/Si interface and thermally activated tunneling processes. The fact that fewer islands than predicted by the orthodox model exhibited SET was also investigated. Schottky barrier lowering due to Fermi level pinning by surface states of the clean silicon surface and a wide depletion region in the substrate were identified as the most likely reasons for the lack of observable SET. The results indicating the genuine nature of the SET suggest that the Schottky barrier may be used as one of the tunnel junctions, and by extension, that islands may be used as a basis for future nanoelectronic devices. However, the lack of observable SET due to Schottky barrier lowering indicates that such devices would be unreliable. To solve the problems of reliability, surface engineering schemes have been proposed to create a tunable tunneling barrier using surface states. Such surface engineering schemes would improve the reliability and flexibility of future nanoelectronic devices.

**Acknowledgements**

The authors would like to thank J.N. Hanson for performing AFM thickness measurements to calibrate the titanium deposition rate and M.C. Zeman for helpful discussions. This work was supported by the National Science Foundation through Grant No. DMR-0512591.

**Figures**

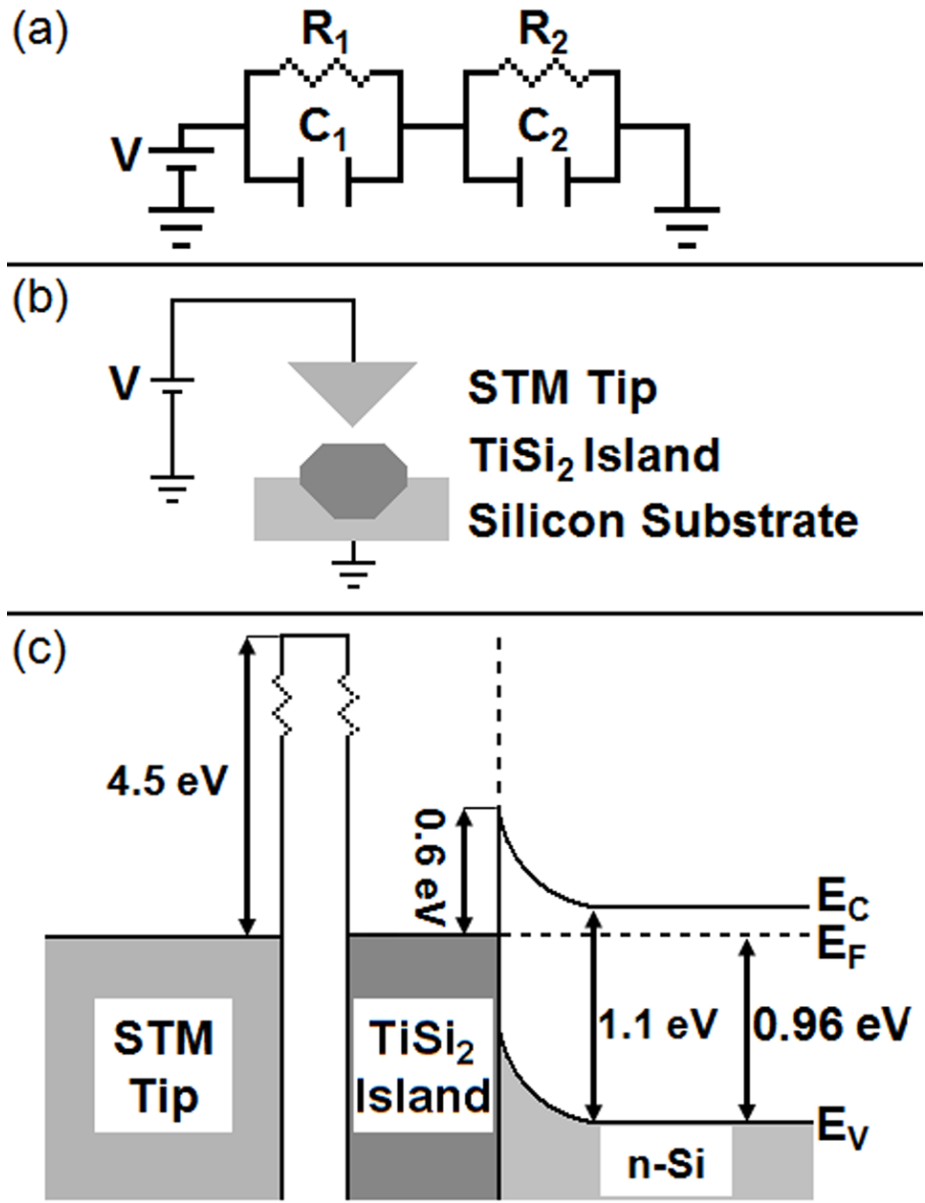

Fig. 1. (a) Schematic representation of a DBTJ. (b) Simplified drawing of the DBTJ structure in these experiments. (c) Diagram of the band structure associated with the DBTJ structure shown in Fig. 1(a). Note that this diagram represents the band structure for the equilibrium case (V=0) at 300 K when $N_D \sim 10^{17}$ cm$^{-3}$.

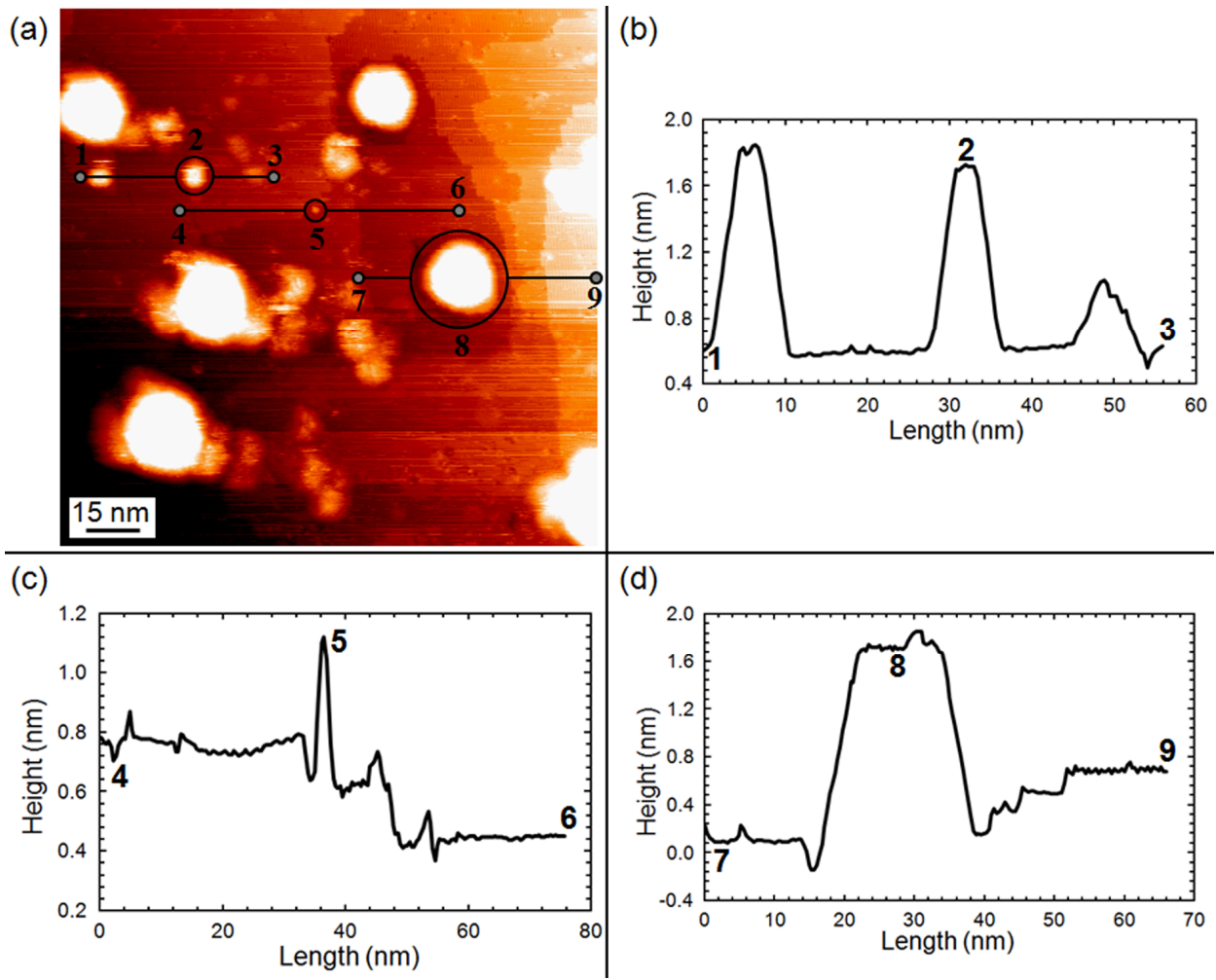

Fig. 2. (a) STM image of TiSi$_2$ islands on n-type Si(100):2×1. Scan size: 150 nm. Tip bias: +1.0 V. Tunneling setpoint: 1.0 nA. (b) Line scan of the island labeled by numbers 1 to 3. (c) Line scan of the island labeled by numbers 4 to 6. (d) Line scan of the island labeled by numbers 7 to 9.

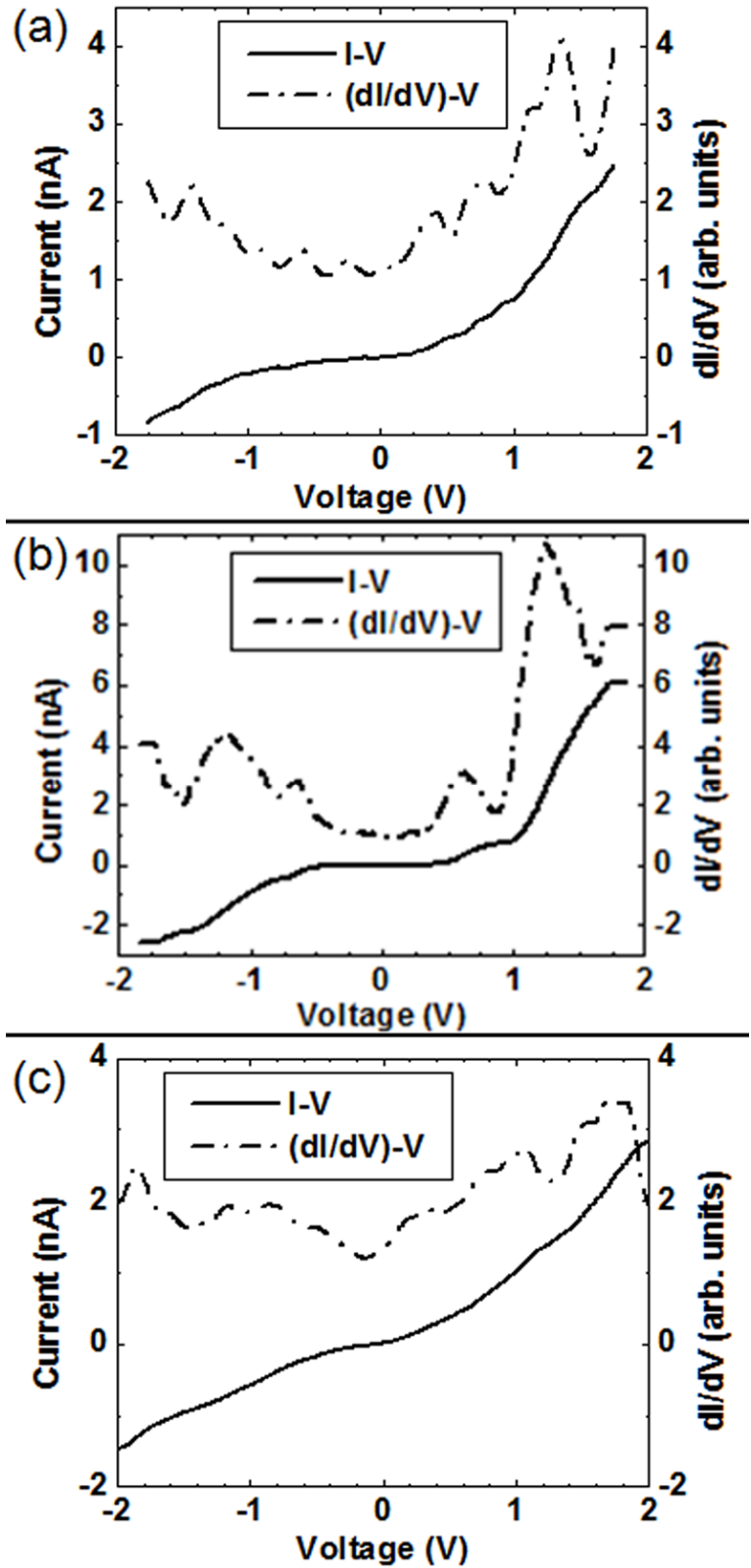

Fig. 3. Current-voltage and (dI/dV)-V curves of the islands in Fig. 2(a) labeled (a) "2" (b) "5," and (c) "8."

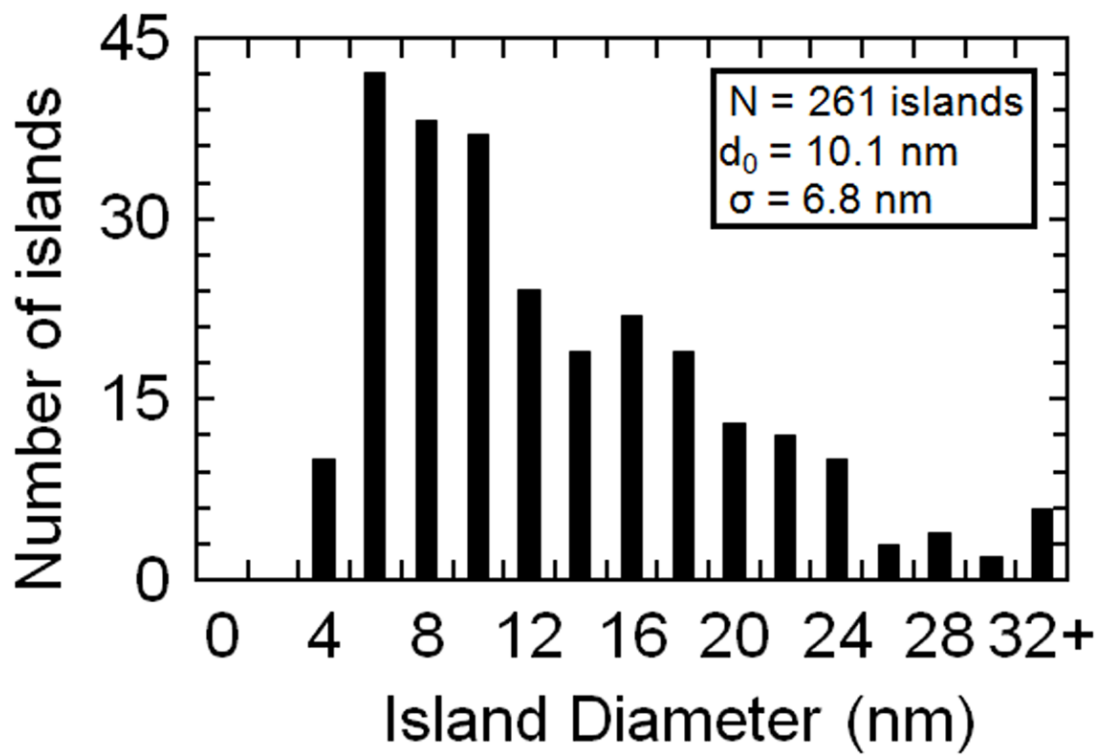

Fig. 4. The distribution of island diameters for round TiSi$_2$ islands in this study from which I-V spectra were recorded.

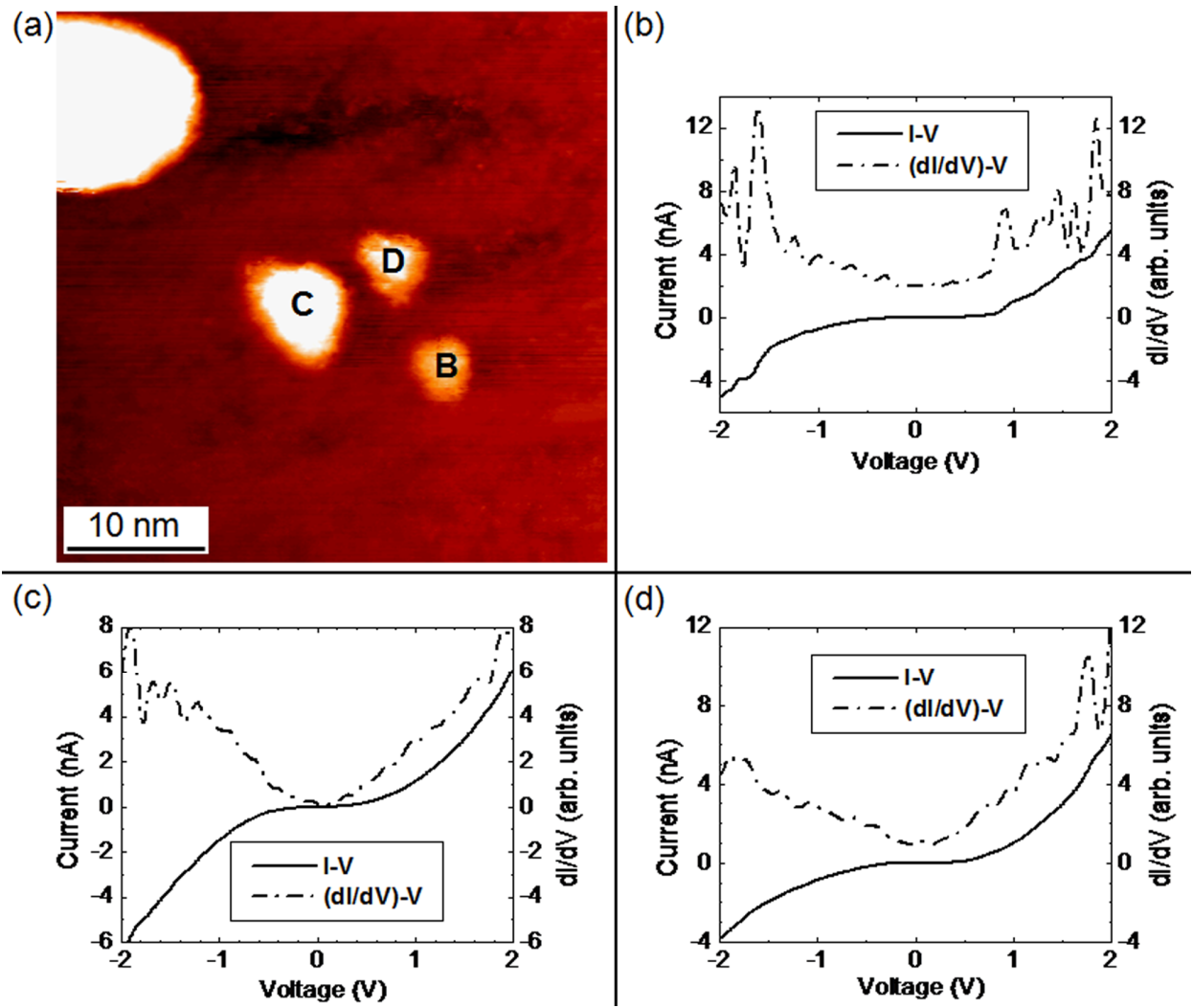

Fig. 5. (a) STM image of TiSi$_2$ islands close together. Scan size: 40 nm. Tip bias: +1.0 V. Tunneling setpoint: 1.0 nA. The distance between islands "B" and "C" is ~2.5 nm while the distance between islands "C" and "D" is ~1.3 nm. Current-voltage and (dI/dV)-V curves recorded from (b) island "B," (c) island "C," and (d) island "D."